\documentclass[prl,floatfix,showpacs,preprintnumbers,10pt,twocolumn]{revtex4}%
\usepackage{amssymb}
\usepackage{amsfonts}
\usepackage{amsmath}
\usepackage{graphicx}
\usepackage{epsfig,float}
\usepackage{dcolumn}
\usepackage{bm}%
\providecommand{\U}[1]{\protect\rule{.1in}{.1in}}
\begin{document}
\title{Drastic disorded-induced reduction of signal amplification in scale-free networks}
\author{Ricardo Chac\'{o}n$^{1}$ and Pedro J. Mart\'{\i}nez$^{2}$}
\affiliation{$^{1}$Departamento de F\'{\i}sica Aplicada, E.I.I., Universidad de
Extremadura, Apartado Postal 382, E-06006 Badajoz, Spain and Instituto de
Computaci\'{o}n Cient\'{\i}fica Avanzada, Universidad de Extremadura, E-06006
Badajoz, Spain}

\affiliation{$^{2}$Departamento de F\'{\i}sica Aplicada, E.I.N.A., Universidad de Zaragoza,
E-50018 Zaragoza, Spain and Instituto de Ciencia de Materiales de Arag\'{o}n,
CSIC-Universidad de Zaragoza, E-50009 Zaragoza, Spain}
\date{\today}

\begin{abstract}
Understanding information transmission across a network is a fundamental task
for controlling and manipulating both biological and man-made information
processing systems. Here, we show how topological resonant-like amplification
effects in scale-free networks of signaling devices are drastically reduced
when phase disorder in the external signals is considered. This is
demonstrated theoretically by means of a star-like network of overdamped
bistable systems, and confirmed numerically by simulations of scale-free
networks of such systems. The taming effect of the phase disorder is found to
be sensitive to the amplification's strength, while the topology-induced
amplification mechanism is robust against this kind of quenched disorder in
the sense that it does not significantly change the values of the coupling
strength where amplification is maximum in its absence.

\end{abstract}

\maketitle

\section{I. INTRODUCTION}
During the last decade, there has been considerable interest in a class of
real-world networks known as scale-free networks [1,2] which have the property
that the degrees, $\kappa$, of the node follow a scale-free power-law
distribution $\left(  P\left(  \kappa\right)  \sim\kappa^{-\gamma},\gamma
\in\lbrack2,3]\right)  $. Examples are social networks such as collaboration
networks, some metabolic and cellular networks, and computer networks such as
the World Wide Web. They exhibit two characteristic properties: robustness
with respect to random failures and fragility with respect to directed attack
[3,4]. Besides topological investigations [5,6], current interest in these
(and other) networks has extended to their controllability [7,8], i.e., to the
characterization and control of the dynamical properties of processes
occurring in them, such as transport [9], synchronization of individual
dynamical behavior occurring at a network's vertices [10,11], role of quenched
spatial disorder in the optimal path problem in weighted networks [12], and
dynamic pattern evolution [13]. Of special relevance is the propagation and
enhancement of resonant collective behaviour across a network due to the
application of weak external signals because of its importance in both
biological and man-made information-processing systems. In this regard, it has
been recently studied the amplification of the response to weak
external signals in networks of
bistable signaling devices [14-17]. In these works, however, the robustness of
the signal amplification against disordered distributions of external signals
was not considered. Clearly, the assumption of homogeneity of the external
signals means that the output is exactly the same for all driving
systems$-$whatever they might be. This mathematically advantageous assumption
(i.e., synchronous driving) is untenable for most of natural and artificial
information-processing systems since a certain amount of randomness is an
unavoidable characteristic of their environments. Thus, to approach signal
amplification phenomena in real-world networks, it seems appropriate to
consider randomness-induced heterogeneous distributions of the external
signals in the model systems.

In this work, we study the interplay between heterogeneous connectivity and
quenched spatial and temporal disorder in random scale-free networks of
signaling devices through the example of a deterministic overdamped bistable
system. This system is sufficiently simple to obtain analytical predictions
while retaining the universal characteristic of a two-state system. The system
reads%
\begin{equation}
\overset{.}{x}_{i}=x_{i}-x_{i}^{3}+\tau\sin\left(  \Omega t+\varphi
_{i}\right)  -\lambda L_{ij}x_{j,}\;i=1,...,N, \tag{1}%
\end{equation}
where $\lambda$ is the coupling, $L_{ij}=\kappa_{i}\delta_{ij}-A_{ij}$ is the
Laplacian matrix of the network, $\kappa_{i}=\sum_{j}A_{ij}$ is the degree of
node $i$, and $A_{ij}$ is the adjacency matrix with entries 1 if $i$ is
connected to $j$, and 0 otherwise. We study the effect of phase disorder on
signal amplification by randomly choosing the initial phases $\varphi_{i}$
uniformly and independently from the interval $\left[  -k\pi,k\pi\right]  $,
with $k\in\left[  0,1\right]  $ being the disorder parameter. Extensive
numerical simulations of the system (1) were conducted for different network
topologies to characterize the amplification-synchronization transition as the
coupling strength is increased. To quantitatively describe this transition, we
used the average amplification $\left\langle \left\langle G\right\rangle
\right\rangle \equiv\max_{i}x_{i}/\tau$ over distinct initial conditions and
phase disorder realizations on one hand, and the synchronization coefficient
[18]
\begin{equation}
\rho=\frac{\left\langle \overline{x_{i}}^{2}\right\rangle -\left\langle
\overline{x_{i}}\right\rangle ^{2}}{\overline{\left\langle x_{i}%
^{2}\right\rangle -\left\langle x_{i}\right\rangle ^{2}}}, \tag{2}%
\end{equation}
on the other hand, where the overlines indicate average over nodes, while the
angle brackets indicate temporal average over a period $T=2\pi/\Omega$.
\section{II. STAR-LIKE NETWORK}
We begin by considering a star-like network of overdamped bistable systems:%
\begin{align}
\overset{.}{x}_{H}  &  =\left[  1-\lambda\left(  N-1\right)  \right]
x_{H}-x_{H}^{3}+\tau\sin\left(  \Omega t+\varphi_{H}\right)  +\lambda
\sum_{i=1}^{N-1}y_{i},\nonumber\\
\overset{.}{y}_{i}  &  =\left(  1-\lambda\right)  y_{i}-y_{i}^{3}+\tau
\sin\left(  \Omega t+\varphi_{i}\right)  +\lambda x_{H}, \tag{3}%
\end{align}
which describes the dynamics of a highly connected node (or hub), $x_{H}$, and
$N-1$ linked systems (or leaves), $y_{i}$. We consider the case of
sufficiently small coupling, $\lambda$, and external signal amplitude, $\tau$,
such that the dynamics of the leaves may be decoupled from that of the hub, on
one hand, and may be suitably described by linearizing their equations around
one of the potential minima, on the other. Thus, one straightforwardly obtains%
\begin{widetext}
\begin{equation}
y_{i}\left(  t\rightarrow\infty\right)  \sim\xi_{i}+\frac{\tau\left[  \left(
2\sin\varphi_{i}-\omega\cos\varphi_{i}\right)  \cos\left(  \omega t\right)
+\left(  \omega\sin\varphi_{i}+2\cos\varphi_{i}\right)  \sin\left(  \omega
t\right)  \right]  }{4+\omega^{2}}, \tag{4}%
\end{equation}
\end{widetext}
where $\xi_{i}=\pm1$ depending on the initial conditions. Since the initial
conditions are randomly chosen, this means that the quantities $\xi_{i}$
behave as discrete random variables governed by Rademacher distributions.
After inserting Eq. (4) into Eq. (3) and solving the resulting equation for
the hub,
\begin{equation}
\overset{.}{x}_{H}=\left[  1-\lambda\left(  N-1\right)  \right]  x_{H}%
-x_{H}^{3}+A^{\prime}\sin\left(  \omega t\right)  +B^{\prime}\cos\left(
\omega t\right)  +\lambda\eta, \tag{5}%
\end{equation}
where
\begin{align*}
\eta &  \equiv\sum_{i=1}^{N-1}\xi_{i},\\
\frac{A^{\prime}}{\tau}  &
\equiv\cos\varphi_{H}+\frac{\lambda\sum_{i=1}^{N-1}  \left(  
\omega\sin\varphi_{i}+2\cos\varphi_{i}\right)  }%
{4+\omega^{2}},\\
\frac{B^{\prime}}{\tau}  &  \equiv\sin\varphi_{H}+\frac{\lambda
\sum_{i=1}^{N-1}\left(  2\sin\varphi_{i}-\omega\cos\varphi_{i}\right)  }%
{4+\omega^{2}},
\end{align*}
one straightforwardly obtains%
\begin{align}
&x_{H}\left(t\rightarrow\infty\right)\sim x_{H}^{(0)}+&\nonumber \\
&\frac{\left(
B^{\prime}\omega-A^{\prime}a_{H}\right)  \sin\left(  \omega t\right)  -\left(
B^{\prime}a_{H}+A^{\prime}\omega\right)  \cos\left(  \omega t\right)  }%
{\omega^{2}+a_{H}^{2}}, \tag{6}%
\end{align}
where $a_{H}\equiv V_{H}^{\prime\prime}\left(  x_{H}^{(0)}\right)  =-\left\{
\frac{3\lambda\eta}{x_{H}^{(0)}}+2\left[  1-\lambda\left(  N-1\right)
\right]  \right\}  $ with $x_{H}^{(0)}$ being the equilibrium in the absence
of external signal while $V_{H}(x_{H})\equiv-\sqrt{h}x_{H}^{2}+x_{H}^{4}/4$ is
the hub potential with $h=\left[  1-\left(  N-1\right)  \lambda\right]
^{2}/4$ being the height of the potential barrier. For finite $N$, the
quantity $\eta$ behaves as a discrete random variable governed by a binomial
distribution with zero mean and variance $N-1$. One sees that the hub's
dynamics is affected by two independent types of quenched disorder: spatial,
through the term $\lambda\eta$, and temporal through the amplitudes
$A^{\prime},B^{\prime}$. For the case of synchronous driving $\left(
\varphi_{i}=\varphi_{H}=0\right)  $, a key observation is that the signal
amplification depends solely on the barrier of the hub potential and the
external signal's amplitude, but \textit{not} on the external signal's sign.
Therefore, for the present case of external signals with phase disorder, the
central limit theorem predicts that the functions $A^{\prime},B^{\prime}$
should be considered as random variables governed by a {\it{folded}} normal (FN)
distribution [19] when $N\rightarrow\infty$ instead of a standard normal
distribution, since the algebraic sign of the external signals plays no role
in the topology-induced signal amplification scenario. For sufficiently large
$N$, this means that one can consider the effective (mean field) equation%
\begin{equation}
\overset{.}{x}_{H}=\left[  1-\lambda\left(  N-1\right)  \right]  x_{H}%
-x_{H}^{3}+A^{\prime\prime}\sin\left(  \omega t\right)  +B^{\prime\prime}%
\cos\left(  \omega t\right)  +\lambda\eta, \tag{7}%
\end{equation}
where
\begin{align*}
\frac{A^{\prime\prime}}{\tau}  &  \equiv\left[  1+\frac{2\lambda\left(
N-1\right)  }{4+\omega^{2}}\right]  \left\langle \cos\varphi_{i}\right\rangle
_{FN}+\frac{\lambda\left(  N-1\right)  \omega\left\langle \sin\varphi
_{i}\right\rangle _{FN}}{4+\omega^{2}},\\
\frac{B^{\prime\prime}}{\tau}  &  \equiv\left[  1+\frac{2\lambda\left(
N-1\right)  }{4+\omega^{2}}\right]  \left\langle \sin\varphi_{i}\right\rangle
_{FN}-\frac{\lambda\left(  N-1\right)  \omega\left\langle \cos\varphi
_{i}\right\rangle _{FN}}{4+\omega^{2}},
\end{align*}
with the averages
\begin{widetext}
\begin{align*}
\left\langle \sin\varphi_{i}\right\rangle _{FN} & \equiv \left\{  \left[
1-\operatorname{sinc}\left(  2k\pi\right)  \right]  /\pi\right\}  ^{1/2},\\
\left\langle \cos\varphi_{i}\right\rangle _{FN}& \equiv \left\{  \left[
1+\operatorname{sinc}\left(  2k\pi\right)  -2\operatorname{sinc}^{2}\left(
k\pi\right)  \right]  /\pi\right\}  ^{1/2} \exp\left\{  -\operatorname{sinc}%
^{2}\left(  k\pi\right)  /\left[  1+\operatorname{sinc}\left(  2k\pi\right)
-2\operatorname{sinc}^{2}\left(  k\pi\right)  \right]  \right\} \\
  &-\operatorname{sinc}(k\pi)\operatorname{erf}\left\{  -\operatorname{sinc}%
\left(  k\pi\right)  /\left[  1+\operatorname{sinc}\left(  2k\pi\right)
-2\operatorname{sinc}^{2}\left(  k\pi\right)  \right]  ^{1/2}\right\}  ,
\end{align*}
\end{widetext}
and where $\operatorname{sinc}(x)\equiv\sin\left(  x\right)  /x$, to reliably
characterize the averaged effect of phase disorder on the topology-induced
signal amplification scenario. Thus, comparing the detailed and effective hub
dynamics equations (Eqs. (5) and (7), respectively), one has that the
effective asymptotic evolution of the hub is given by Eq. (6) with the
substitutions $A^{\prime}\rightarrow A^{\prime\prime},B^{\prime}\rightarrow
B^{\prime\prime},$ and hence%
\begin{equation}
G_{eff}\left(  \eta\right)  =\frac{\sqrt{\left(  B^{\prime\prime}%
\omega-A^{\prime\prime}a_{H}\right)  ^{2}+\left(  B^{\prime\prime}%
a_{H}+A^{\prime\prime}\omega\right)  ^{2}}}{\tau\left(  \omega^{2}+a_{H}%
^{2}\right)  } \tag{8}%
\end{equation}
provides an estimate of its amplification. For sufficiently large $N$, we may
assume that the quantity $\eta$ behaves as a continuous random variable
governed by a standard normal distribution, and hence%
\begin{equation}
\left\langle G_{eff}\right\rangle =\frac{1}{\sqrt{2\pi\left(  N-1\right)  }%
}\int_{-\infty}^{\infty}G_{eff}\left(  \eta\right)  \exp\left[  \frac
{-\eta^{2}}{2\left(  N-1\right)  }\right]  d\eta\tag{9}%
\end{equation}
provides the final average amplification. Equation (9) predicts that
$\left\langle G_{eff}\right\rangle \left(  \lambda,N,\omega,k>0\right)
<\left\langle G_{eff}\right\rangle \left(  \lambda,N,\omega,k=0\right)  $ and
that the signal amplification \textit{decreases monotonously} on average as
the strength of the phase disorder is increased (i.e., as $k$ is increased;
see Fig. 1, left panel), which is accurately confirmed by numerical simulations
(cf. Fig. 1, right panel). One also has from Eq. (9) that $\left\langle
G_{eff}\right\rangle \left(  \lambda,N,\omega,k\right)  $, as a function of
only $\lambda$, presents a sharp single maximum at $\lambda\approx\left(
N-1\right)  ^{-1}$for all $k$, which indicates that the topology-induced
amplification mechanism is \textit{robust} against phase disorder in star-like networks.

\begin{figure}[htb]
\epsfig{file=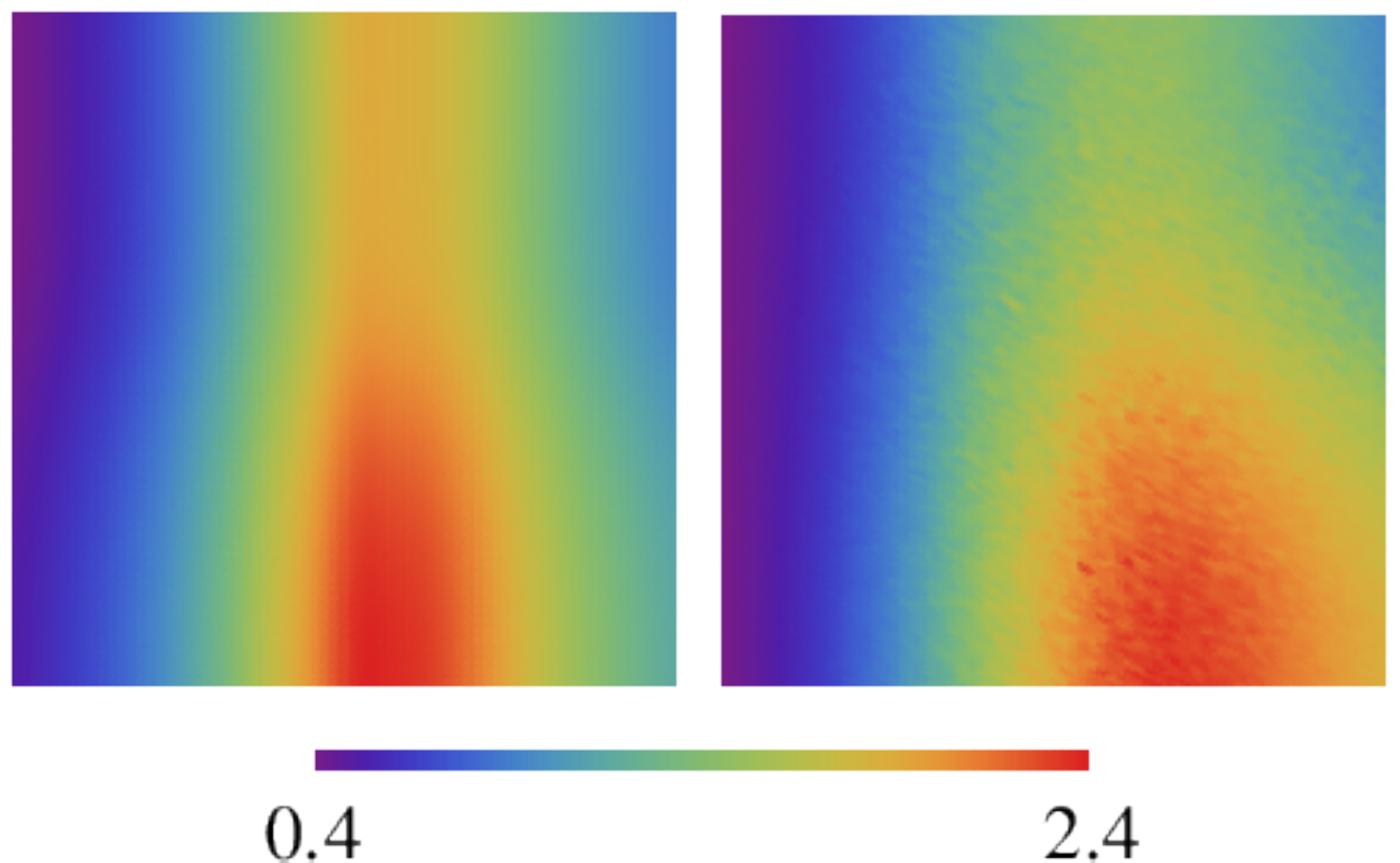,width=0.45\textwidth}
\caption{Theoretical average amplification $\left\langle G_{eff}\right\rangle $
in the $\left(  k-\lambda\right)  $ parameter plane with $\lambda \in
\left [ 0,0.0035\right ]$ and $k \in \left [ 0,1\right ]$ (left panel, Eq. (9)) and
corresponding numerical results $\left\langle \left\langle G\right\rangle
\right\rangle $ (right panel) for a starlike network (cf. Eq. (3)) and
$N=500,\omega=2\pi\times10^{-1},\tau=0.01$.}
\end{figure}
\section{III. BARAB\'ASI-ALBERT NETWORK}
Next, we discuss the possibility of extending the results obtained for a
star-like network to Barab\'{a}si-Albert (BA) networks [2] of the same
overdamped bistable systems. Indeed, a highly connected node in the BA network
can be thought of as a hub of a local star-like network with a certain degree
$\kappa$ picked up from the degree distribution. Thus, one can expect that the
suppressory effect of phase disorder will act at any scale yielding a drastic
reduction of the signal amplification over the whole scale-free network.
Figure 2 shows an illustrative example where the averaged amplification
$\left\langle \left\langle G\right\rangle \right\rangle $ is plotted versus
coupling $\lambda$ (top panel) and phase disorder parameter $k$ (bottom
panel). 
\begin{figure}[htb]
\epsfig{file=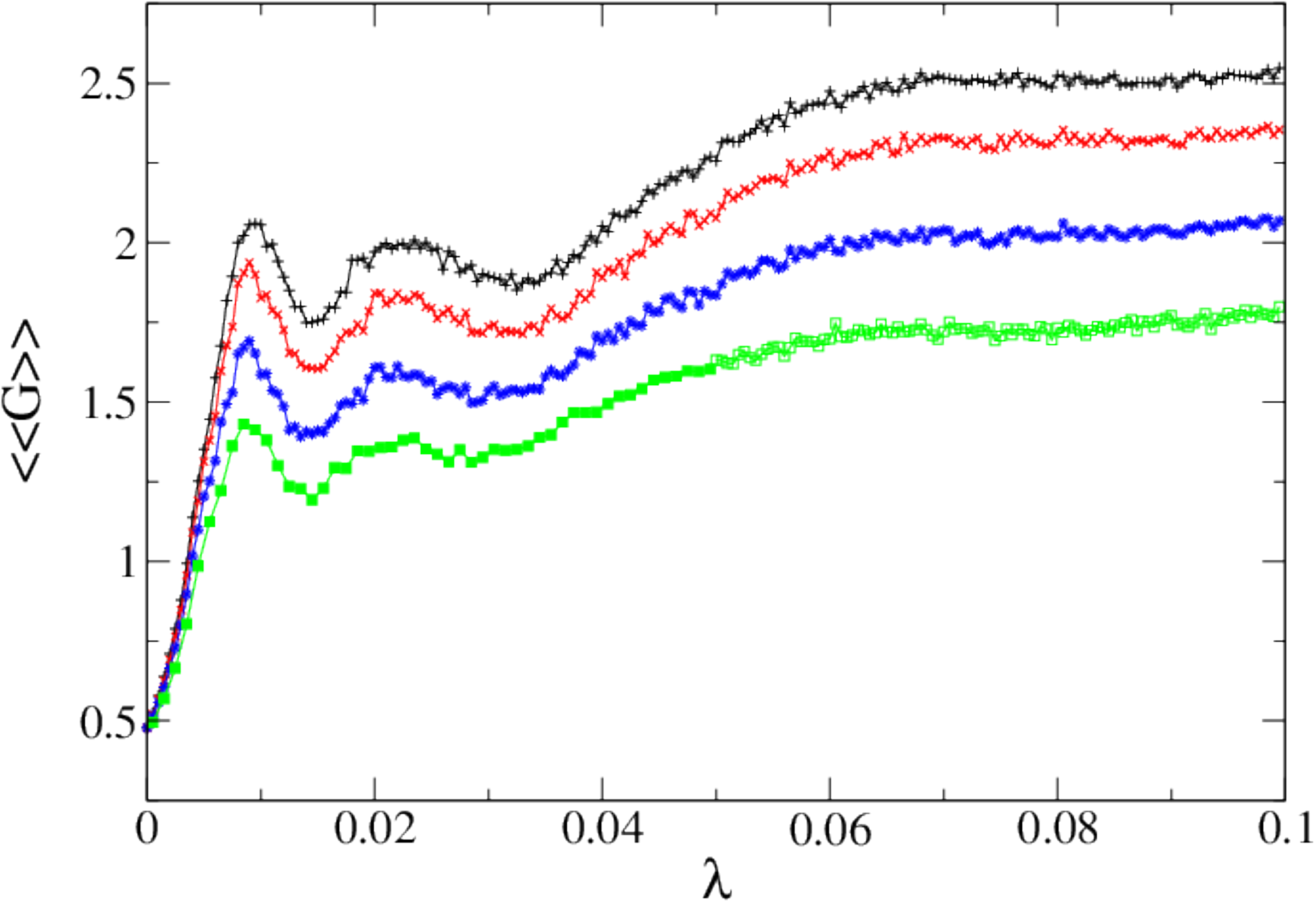,width=0.45\textwidth}
\epsfig{file=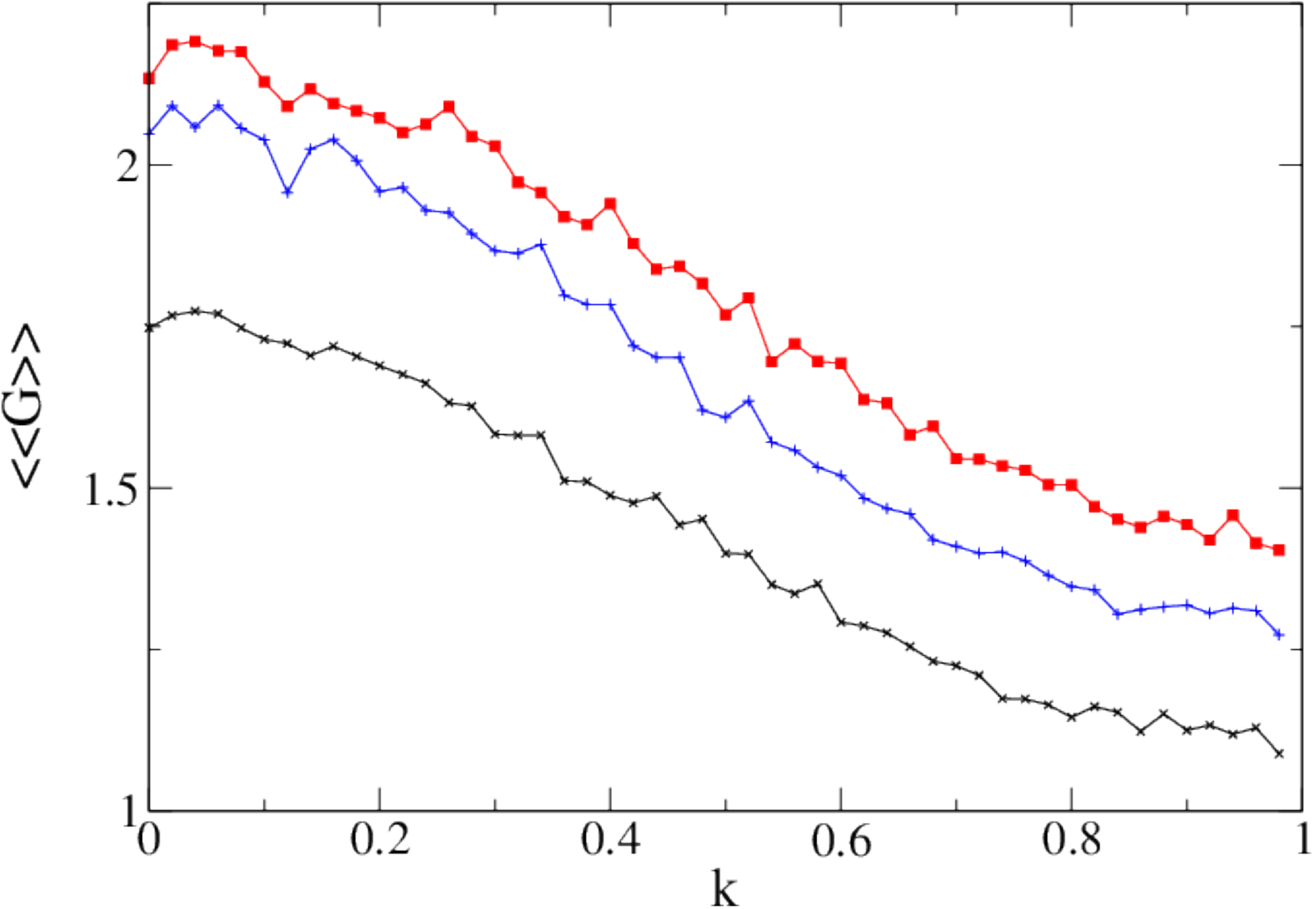,width=0.45\textwidth}
\caption{Top panel: Average amplification $\left\langle \left\langle
G\right\rangle \right\rangle $ versus coupling $\lambda$ for a BA scale-free
network and four values of the phase disorder parameter: $k=0\ (+)$,
$k=0.3\ (\times)$, $k=0.5$ (stars), and $k=0.7$ (squares). Note that the first
relative maximum of $\left\langle \left\langle G\right\rangle \right\rangle $
occurs around $\lambda\approx0.008$ for the four values of $k$ while the
network has a maximal active hub having $136$ leaves. For this effective
star-like network the theoretically predicted maximum occurs at $\lambda
=\lambda_{\max,1}\approx0.0074$. Bottom panel: Average amplification
$\left\langle \left\langle G\right\rangle \right\rangle $ versus phase
disorder parameter $k$ for a BA scale-free network and three values of the
coupling: $\lambda=0.009\ \left(  +\right)  $, $\lambda=0.015\ (\times)$, and
$\lambda=0.045$ (squares). Averaged degree $\left\langle \kappa\right\rangle
=3$, $\gamma=2.7$, and the remaining fixed parameters are as in Fig. 1.}
\end{figure}

One sees that $\left\langle \left\langle G\right\rangle \right\rangle
$ becomes ever smaller as $k$ increases over the complete range of values of
$\lambda$, confirming the predictions of the above theoretical analysis. As
the coupling $\lambda$ is increased from 0, an increasing number of effective
star-like networks embedded in the scale-free network become active in the
sense that their hubs are the only nodes undertaking a significative
amplification of their responses on average. This is the weak coupling regime
where the scale-free network's dynamics can therefore be understood from that
of a star-like network. Also, the first relative maximum of $\left\langle
\left\langle G\right\rangle \right\rangle $ of the scale-free network as a
function of coupling $\lambda$ was systematically found at the value
$\lambda=\lambda_{\max,1}$ predicted from the star-like network analysis for
the only active hub (the most connected) existing at $\lambda=\lambda_{\max
,1}$ (see Fig. 2, top panel). By increasing slightly $\lambda$ from
$\lambda_{\max,1}$ yields the additional activation of the second most
connected node such that there are now two effective star-like networks which
have, for $N$ sufficiently large, a high probability of being isolated each
other. Since the averaged amplification of a star-like network exhibits a
single maximum as a function of the coupling which is very sharp (cf. Eq.
(9)), when $\lambda\gtrsim\lambda_{\max,1}$ the averaged amplification of the
most connected hub drastically decreases with respect to its value at
$\lambda=\lambda_{\max,1}$, while the averaged amplification of the second
most connected hub should also be relatively small owing to its lower number
of leaves. This explains the existence of the aforementioned first relative
maximum at $\lambda=\lambda_{\max,1}$ since the averaged amplification of the
scale-free network is no more than the sum of the averaged amplifications of
the most connected hubs provided that $N$ is sufficiently large. Further
increase of $\lambda$ yields the activation of additional subsequent most
connected nodes resulting in an increasing value of the averaged amplification
since these hubs still remain unconnected each other (note the appearance of a
secondary relative maximum at a value $\lambda=\lambda_{\max,2}$ irrespective
of the value of $k$, cf. Fig. 2, top panel). 

\begin{figure}[hbt]
\epsfig{file=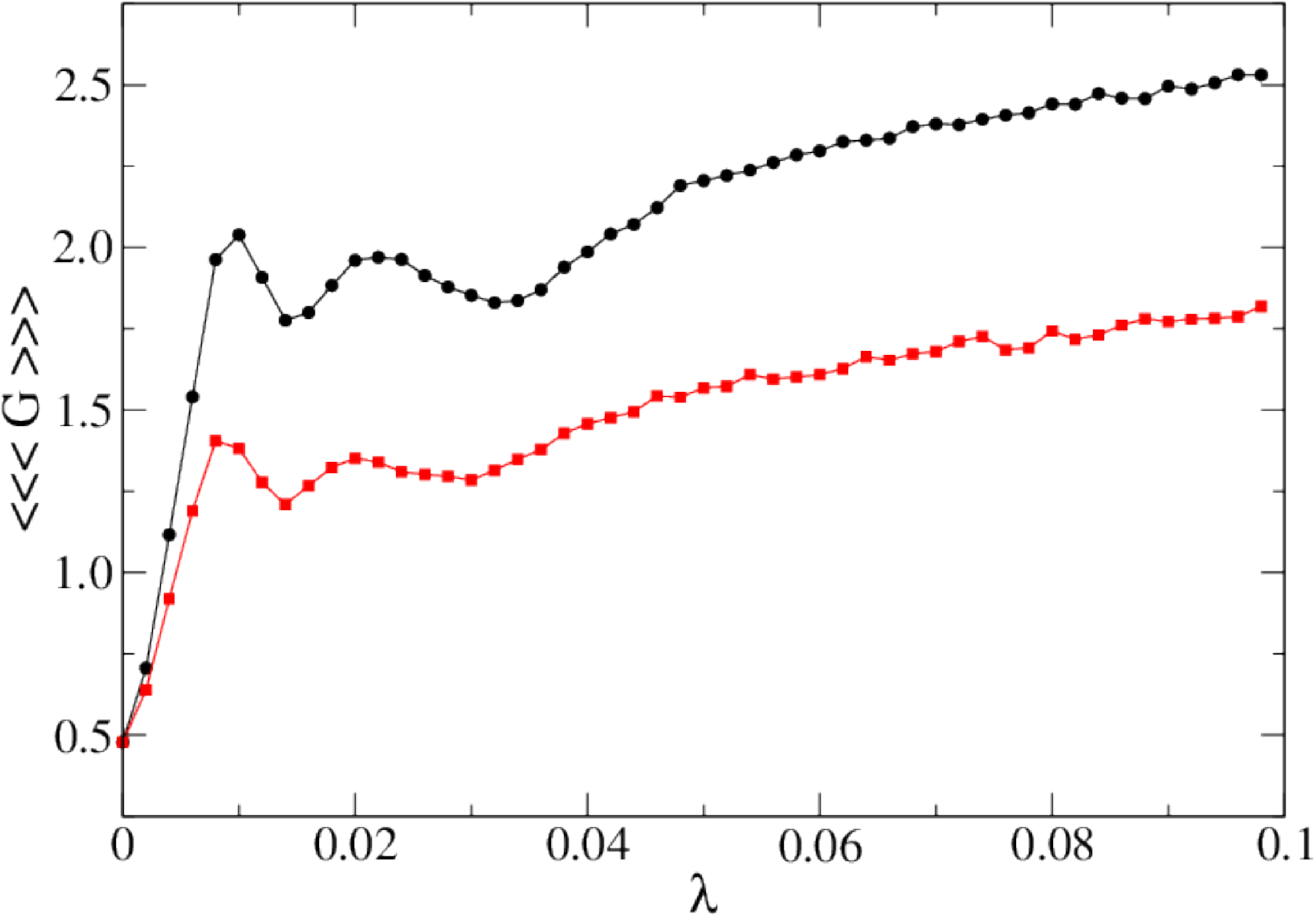,width=0.45\textwidth}
\epsfig{file=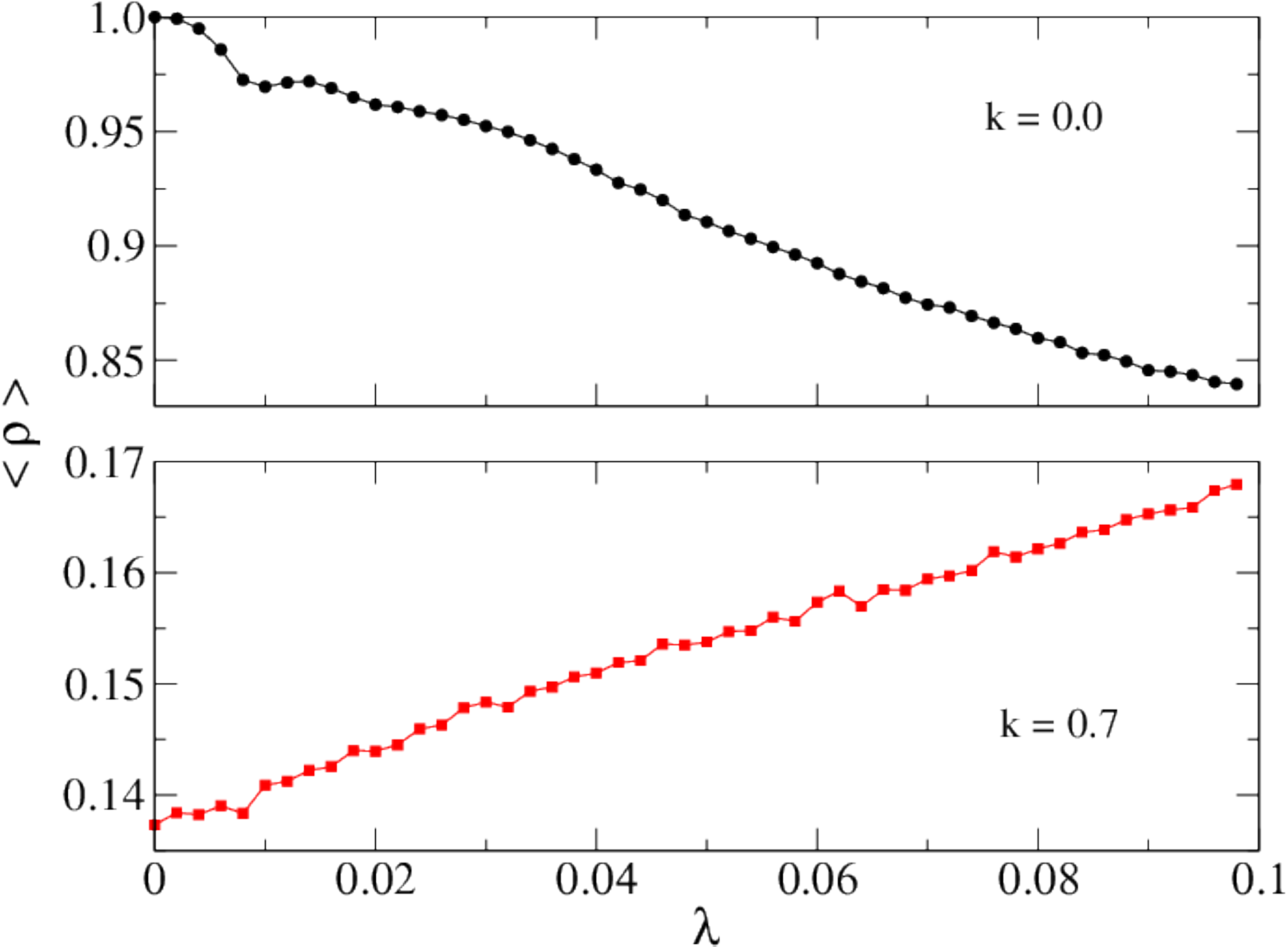,width=0.45\textwidth}
\caption{Average of the average amplification over $10^{2}$ random realizations
of the network connectivity $\left\langle \left\langle \left\langle
G\right\rangle \right\rangle \right\rangle \equiv\left\langle \max_{i}%
x_{i}/\tau\right\rangle $ for two values of the phase disorder parameter
($k=0\ $(circles) and $k=0.7\ $(squares)) (top panel) and corresponding
synchronization coefficient $\left\langle \rho\right\rangle $ (cf. Eq. (2))
for $k=0$ and $k=0.7$ (bottom panels) versus coupling $\lambda$ for a BA
scale-free network with $\gamma=2.7$ and $\left\langle
\kappa\right\rangle =3$. Other fixed parameters are as in Fig. 1.
}
\end{figure}

We found that these two first relative
maxima appear at (approximately) the same values of $\lambda$ in any random
realization of the network connectivity and for any value of $k$ (see
Fig. 3,
top panel). This robustness of the amplification scenario against the presence of
phase disorder does not hold for the synchronization scenario in the weak
coupling regime in the sense that the synchronization monotonously decreases
(increases) as $\lambda$ is increased from 0 in the absence (presence) of
phase disorder (see Fig. 3, bottom panel). This can be understood as the result of
two cojoint mechanisms: the disorder-induced lowering of amplification and the
coupling-induced increasing of synchronization. Indeed, in the absence of the
former mechanism $\left(  k=0\right)  $, the latter mechanism by itself is not
enough to dominate the desynchronization effect of the topology-induced
amplification mechanism. We additionally found that, for \textit{any} value of
the phase disorder parameter $k\geqslant0$, the averaged amplification
(synchronization) decreases (increases) as the power-law distribution exponent
$\gamma$ is increased (see Fig. 4), providing thus an additional confirmation
of the robustness of the topology-induced amplification mechanism against the
presence of phase disorder.
\begin{figure}[htb]
\epsfig{file=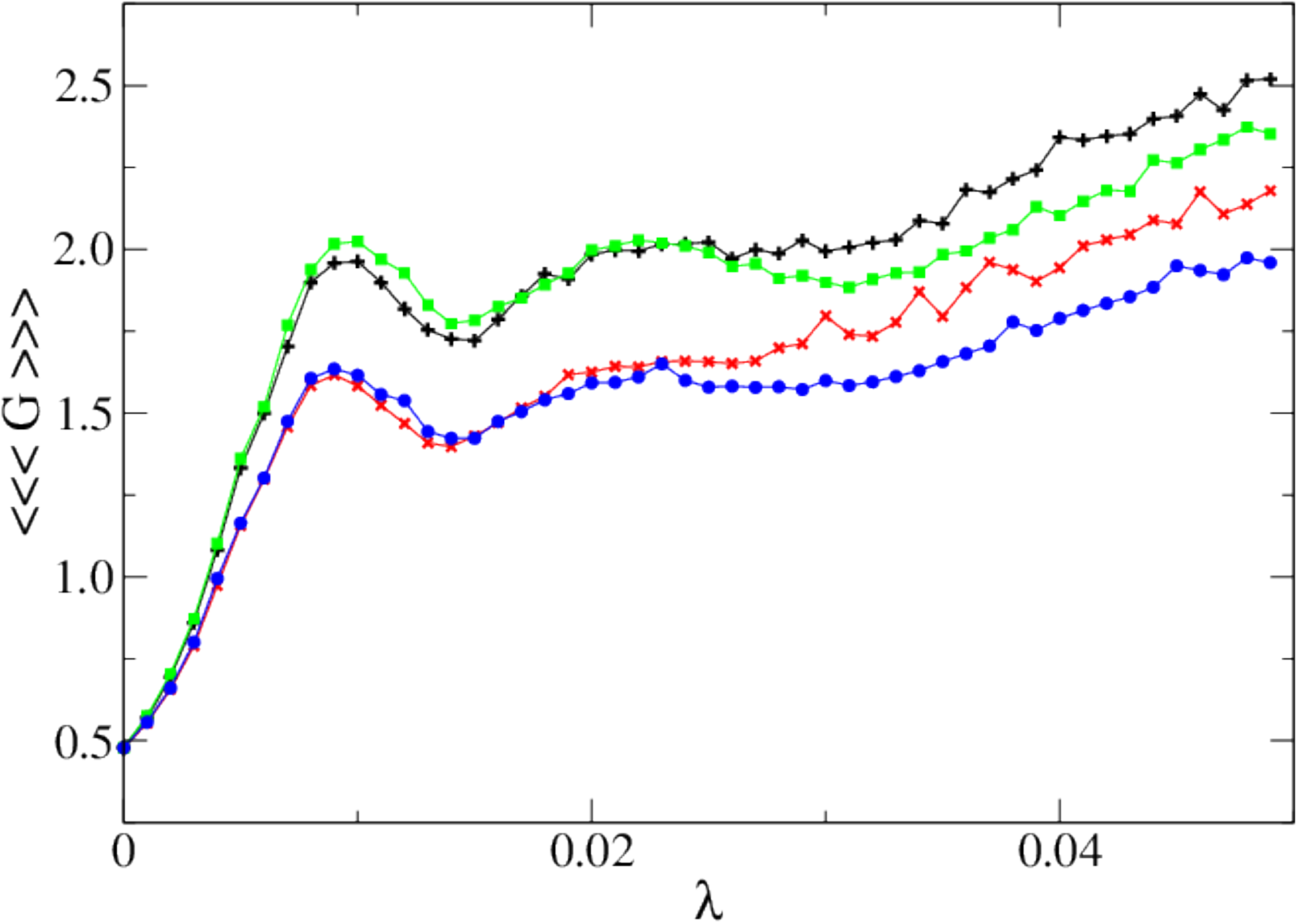,width=0.45\textwidth}
\epsfig{file=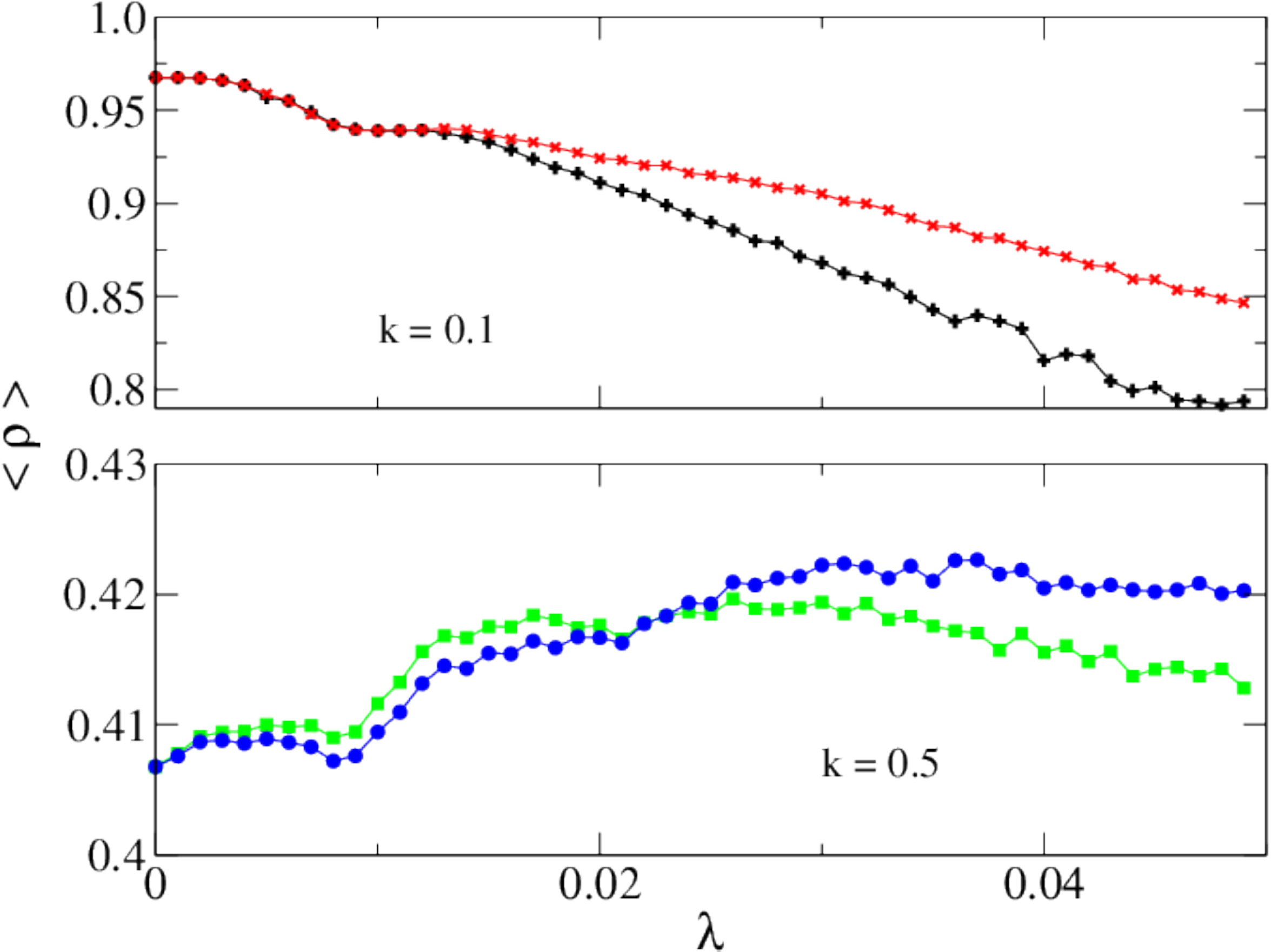,width=0.45\textwidth}
\caption{Average of the average amplification over $30$ random realizations
of the network connectivity $\left\langle \left\langle \left\langle
G\right\rangle \right\rangle \right\rangle \equiv\left\langle \max_{i}%
x_{i}/\tau\right\rangle $ (top panel) and corresponding synchronization
coefficient $\left\langle \rho\right\rangle $ (cf. Eq. (2), medium and bottom
panels) versus coupling $\lambda$ for a BA scale-free network with
$\left\langle \kappa\right\rangle =3$ and different
values of the phase disorder parameter and the power-law distribution
exponent: $(k,\gamma)=(0.1,2)$ ($+$), $\left(  0.1,2.5\right)  $ ($\times$),
$(0.5,2)$ (squares), $\left(  0.5,2.5\right)  $ (circles). Other fixed
parameters are as in Fig. 1.
}
\end{figure}

Figure 5 provides an additional example for a higher average degree
$\left(
\left\langle \kappa\right\rangle =5\right)  $ confirming the above
amplification-synchronization scenario. Also, the range of the weak
coupling
regime where a noticeable amplification occurs diminishes as the average
degree is increased irrespective of the strength of phase disorder
(see Fig. 5).

\begin{figure}[htb]
\epsfig{file=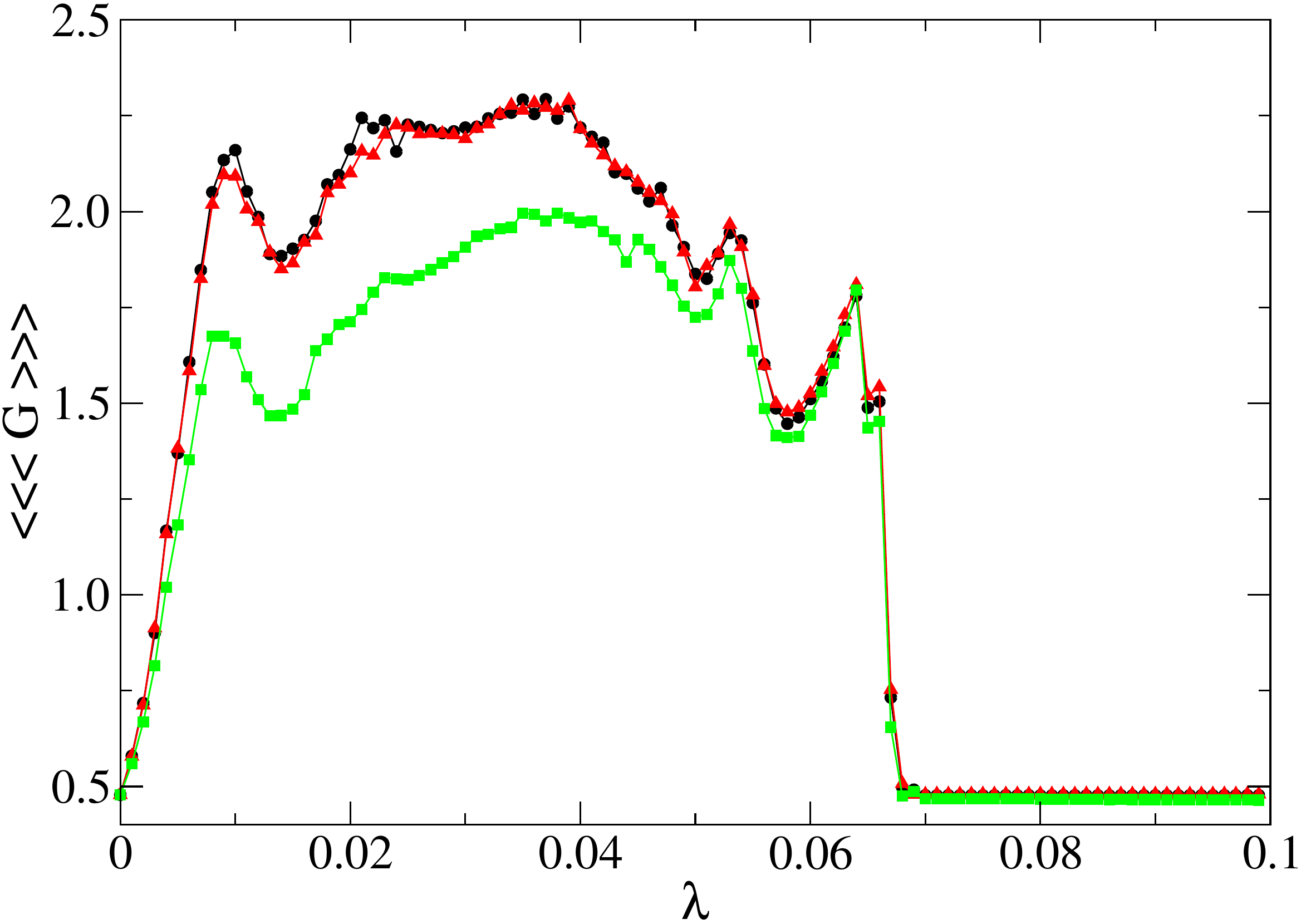,width=0.45\textwidth}
\epsfig{file=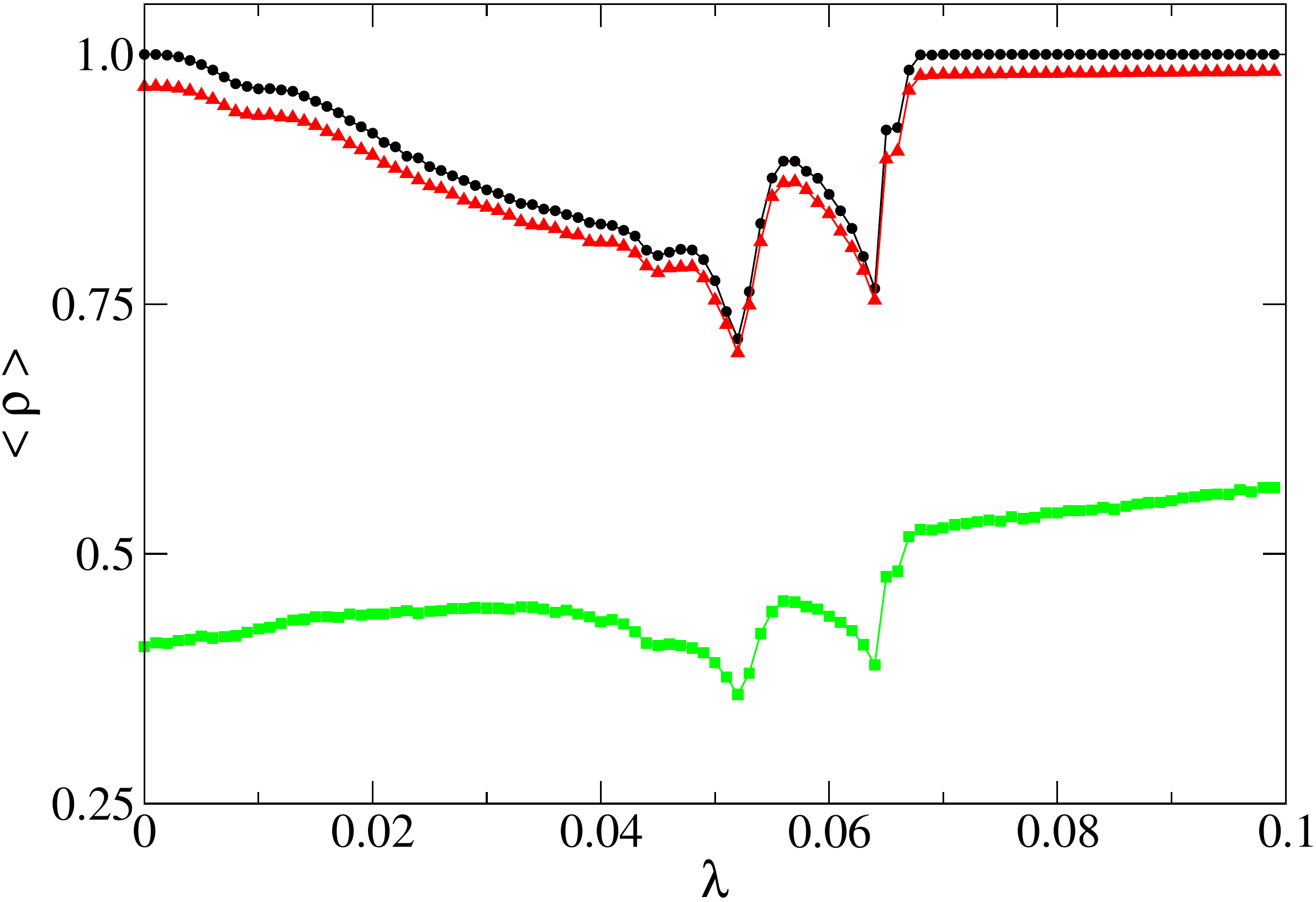,width=0.45\textwidth}
\caption{(Color online) Average of the average amplification over 30
random
realizations of the network connectivity $\left\langle \left\langle
\left\langle G\right\rangle \right\rangle \right\rangle
\equiv\left\langle
\max_{i}x_{i}/\tau\right\rangle $ (top panel) and corresponding
synchronization coefficient $\left\langle \rho\right\rangle $ (cf. Eq.
(2),
bottom panel) versus coupling $\lambda$ for a BA scale-free network with
$\left\langle \kappa\right\rangle =5,\gamma=2.7$, and three values of
the
phase disorder parameter: $k=0$ (circles), $0.1$ (triangles), $0.5$
(squares).
Other fixed parameters are as in Fig. 1.
}
\end{figure}

\section{IV. CONCLUSION}
In sum, we have shown through the example of a network of overdamped bistable
systems that phase disorder in the external signals strongly reduces
topology-induced signal amplification in scale-free networks. We have
analytically demonstrated that this effect of quenched temporal disorder may
be completely characterized in the simple case of a starlike network. The
relevance of the present results stems from the fact that phase disorder in
the external signals, contrary to the effect of additive Gaussian
white noise [20] and contrary to what happens in regular networks [21,22] of
chaotic nonautonomous oscillators where phase disorder acts favouring
signal-induced regularization, has a negative effect in the amplification process of external
signals, favouring thus synchronization in scale-free networks. Interestingly,
our results indicate that the presence of phase disorder does not
significantly change the values of the coupling strength where amplification
is maximum in its absence (i.e., when all nodes are synchronously driven),
which means that the topology-induced amplification mechanism is robust
against this kind of quenched disorder. One is thus tempted to speculate that
this robustness might well provide another reason for the prevalence of
scale-free networks in nature.

We wish to express our gratitude to Prof. J. G\'omez-Garde\~nes for
helpful discussions. 
R. C. and P. J. M. acknowledge financial support from the Ministerio de
Econom\'{\i}a y Competitividad (MECC, Spain) through FIS2012-34902 and
FIS2011-25167 projects, respectively. R. C. acknowledges financial support
from the Junta de Extremadura (JEx, Spain) through project GR10045.

\end{document}